\documentclass[%
reprint,
superscriptaddress,
showpacs,
amsmath,amssymb,
aps,
]{revtex4-1}

\usepackage{graphicx}
\usepackage{dcolumn}
\usepackage{bm}
\usepackage{amsmath}
\usepackage{autobreak}
\usepackage{hyperref}
\usepackage[mathlines]{lineno}
\usepackage{natbib}
\usepackage{multirow}
\usepackage{booktabs}
\usepackage{xspace}
\usepackage{float}
\usepackage{diagbox}

\begin{document}

\title{Search for boosted keV-MeV light dark matter particles from evaporating\\ primordial black holes at the CDEX-10 experiment}

\author{Z.~H.~Zhang}
\affiliation{Key Laboratory of Particle and Radiation Imaging (Ministry of Education) and Department of Engineering Physics, Tsinghua University, Beijing 100084}
\author{L.~T.~Yang}
\email{Corresponding author: yanglt@mail.tsinghua.edu.cn}
\affiliation{Key Laboratory of Particle and Radiation Imaging (Ministry of Education) and Department of Engineering Physics, Tsinghua University, Beijing 100084}
\author{Q.~Yue}
\email{Corresponding author: yueq@mail.tsinghua.edu.cn}
\affiliation{Key Laboratory of Particle and Radiation Imaging (Ministry of Education) and Department of Engineering Physics, Tsinghua University, Beijing 100084}
\author{K.~J.~Kang}
\affiliation{Key Laboratory of Particle and Radiation Imaging (Ministry of Education) and Department of Engineering Physics, Tsinghua University, Beijing 100084}
\author{Y.~J.~Li}
\affiliation{Key Laboratory of Particle and Radiation Imaging (Ministry of Education) and Department of Engineering Physics, Tsinghua University, Beijing 100084}

\author{H.~P.~An}
\affiliation{Key Laboratory of Particle and Radiation Imaging (Ministry of Education) and Department of Engineering Physics, Tsinghua University, Beijing 100084}
\affiliation{Department of Physics, Tsinghua University, Beijing 100084}

\author{Greeshma~C.}
\altaffiliation{Participating as a member of TEXONO Collaboration}
\affiliation{Institute of Physics, Academia Sinica, Taipei 11529}

\author{J.~P.~Chang}
\affiliation{NUCTECH Company, Beijing 100084}

\author{Y.~H.~Chen}
\affiliation{YaLong River Hydropower Development Company, Chengdu 610051}
\author{J.~P.~Cheng}
\affiliation{Key Laboratory of Particle and Radiation Imaging (Ministry of Education) and Department of Engineering Physics, Tsinghua University, Beijing 100084}
\affiliation{College of Nuclear Science and Technology, Beijing Normal University, Beijing 100875}
\author{W.~H.~Dai}
\affiliation{Key Laboratory of Particle and Radiation Imaging (Ministry of Education) and Department of Engineering Physics, Tsinghua University, Beijing 100084}
\author{Z.~Deng}
\affiliation{Key Laboratory of Particle and Radiation Imaging (Ministry of Education) and Department of Engineering Physics, Tsinghua University, Beijing 100084}
\author{C.~H.~Fang}
\affiliation{College of Physics, Sichuan University, Chengdu 610065}
\author{X.~P.~Geng}
\affiliation{Key Laboratory of Particle and Radiation Imaging (Ministry of Education) and Department of Engineering Physics, Tsinghua University, Beijing 100084}
\author{H.~Gong}
\affiliation{Key Laboratory of Particle and Radiation Imaging (Ministry of Education) and Department of Engineering Physics, Tsinghua University, Beijing 100084}
\author{Q.~J.~Guo}
\affiliation{School of Physics, Peking University, Beijing 100871}
\author{X.~Y.~Guo}
\affiliation{YaLong River Hydropower Development Company, Chengdu 610051}
\author{L.~He}
\affiliation{NUCTECH Company, Beijing 100084}
\author{S.~M.~He}
\affiliation{YaLong River Hydropower Development Company, Chengdu 610051}
\author{J.~W.~Hu}
\affiliation{Key Laboratory of Particle and Radiation Imaging (Ministry of Education) and Department of Engineering Physics, Tsinghua University, Beijing 100084}
\author{H.~X.~Huang}
\affiliation{Department of Nuclear Physics, China Institute of Atomic Energy, Beijing 102413}
\author{T.~C.~Huang}
\affiliation{Sino-French Institute of Nuclear and Technology, Sun Yat-sen University, Zhuhai 519082}
\author{H.~T.~Jia}
\affiliation{College of Physics, Sichuan University, Chengdu 610065}
\author{X.~Jiang}
\affiliation{College of Physics, Sichuan University, Chengdu 610065}

\author{S.~Karmakar}
\altaffiliation{Participating as a member of TEXONO Collaboration}
\affiliation{Institute of Physics, Academia Sinica, Taipei 11529}

\author{H.~B.~Li}
\altaffiliation{Participating as a member of TEXONO Collaboration}
\affiliation{Institute of Physics, Academia Sinica, Taipei 11529}
\author{J.~M.~Li}
\affiliation{Key Laboratory of Particle and Radiation Imaging (Ministry of Education) and Department of Engineering Physics, Tsinghua University, Beijing 100084}
\author{J.~Li}
\affiliation{Key Laboratory of Particle and Radiation Imaging (Ministry of Education) and Department of Engineering Physics, Tsinghua University, Beijing 100084}
\author{Q.~Y.~Li}
\affiliation{College of Physics, Sichuan University, Chengdu 610065}
\author{R.~M.~J.~Li}
\affiliation{College of Physics, Sichuan University, Chengdu 610065}
\author{X.~Q.~Li}
\affiliation{School of Physics, Nankai University, Tianjin 300071}
\author{Y.~L.~Li}
\affiliation{Key Laboratory of Particle and Radiation Imaging (Ministry of Education) and Department of Engineering Physics, Tsinghua University, Beijing 100084}
\author{Y.~F.~Liang}
\affiliation{Key Laboratory of Particle and Radiation Imaging (Ministry of Education) and Department of Engineering Physics, Tsinghua University, Beijing 100084}
\author{B.~Liao}
\affiliation{College of Nuclear Science and Technology, Beijing Normal University, Beijing 100875}
\author{F.~K.~Lin}
\altaffiliation{Participating as a member of TEXONO Collaboration}
\affiliation{Institute of Physics, Academia Sinica, Taipei 11529}
\author{S.~T.~Lin}
\affiliation{College of Physics, Sichuan University, Chengdu 610065}
\author{J.~X.~Liu}
\affiliation{Key Laboratory of Particle and Radiation Imaging (Ministry of Education) and Department of Engineering Physics, Tsinghua University, Beijing 100084}
\author{S.~K.~Liu}
\affiliation{College of Physics, Sichuan University, Chengdu 610065}
\author{Y.~D.~Liu}
\affiliation{College of Nuclear Science and Technology, Beijing Normal University, Beijing 100875}
\author{Y.~Liu}
\affiliation{College of Physics, Sichuan University, Chengdu 610065}
\author{Y.~Y.~Liu}
\affiliation{College of Nuclear Science and Technology, Beijing Normal University, Beijing 100875}
\author{Z.~Z.~Liu}
\affiliation{Key Laboratory of Particle and Radiation Imaging (Ministry of Education) and Department of Engineering Physics, Tsinghua University, Beijing 100084}
\author{H.~Ma}
\affiliation{Key Laboratory of Particle and Radiation Imaging (Ministry of Education) and Department of Engineering Physics, Tsinghua University, Beijing 100084}
\author{Y.~C.~Mao}
\affiliation{School of Physics, Peking University, Beijing 100871}
\author{Q.~Y.~Nie}
\affiliation{Key Laboratory of Particle and Radiation Imaging (Ministry of Education) and Department of Engineering Physics, Tsinghua University, Beijing 100084}
\author{J.~H.~Ning}
\affiliation{YaLong River Hydropower Development Company, Chengdu 610051}
\author{H.~Pan}
\affiliation{NUCTECH Company, Beijing 100084}
\author{N.~C.~Qi}
\affiliation{YaLong River Hydropower Development Company, Chengdu 610051}
\author{J.~Ren}
\affiliation{Department of Nuclear Physics, China Institute of Atomic Energy, Beijing 102413}
\author{X.~C.~Ruan}
\affiliation{Department of Nuclear Physics, China Institute of Atomic Energy, Beijing 102413}

\author{Z.~She}
\affiliation{Key Laboratory of Particle and Radiation Imaging (Ministry of Education) and Department of Engineering Physics, Tsinghua University, Beijing 100084}
\author{M.~K.~Singh}
\altaffiliation{Participating as a member of TEXONO Collaboration}
\affiliation{Institute of Physics, Academia Sinica, Taipei 11529}
\affiliation{Department of Physics, Banaras Hindu University, Varanasi 221005}
\author{T.~X.~Sun}
\affiliation{College of Nuclear Science and Technology, Beijing Normal University, Beijing 100875}
\author{C.~J.~Tang}
\affiliation{College of Physics, Sichuan University, Chengdu 610065}
\author{W.~Y.~Tang}
\affiliation{Key Laboratory of Particle and Radiation Imaging (Ministry of Education) and Department of Engineering Physics, Tsinghua University, Beijing 100084}
\author{Y.~Tian}
\affiliation{Key Laboratory of Particle and Radiation Imaging (Ministry of Education) and Department of Engineering Physics, Tsinghua University, Beijing 100084}
\author{G.~F.~Wang}
\affiliation{College of Nuclear Science and Technology, Beijing Normal University, Beijing 100875}
\author{L.~Wang}
\affiliation{Department of  Physics, Beijing Normal University, Beijing 100875}
\author{Q.~Wang}
\affiliation{Key Laboratory of Particle and Radiation Imaging (Ministry of Education) and Department of Engineering Physics, Tsinghua University, Beijing 100084}
\affiliation{Department of Physics, Tsinghua University, Beijing 100084}
\author{Y.~F.~Wang}
\affiliation{Key Laboratory of Particle and Radiation Imaging (Ministry of Education) and Department of Engineering Physics, Tsinghua University, Beijing 100084}
\author{Y.~X.~Wang}
\affiliation{School of Physics, Peking University, Beijing 100871}
\author{H.~T.~Wong}
\altaffiliation{Participating as a member of TEXONO Collaboration}
\affiliation{Institute of Physics, Academia Sinica, Taipei 11529}
\author{S.~Y.~Wu}
\affiliation{YaLong River Hydropower Development Company, Chengdu 610051}
\author{Y.~C.~Wu}
\affiliation{Key Laboratory of Particle and Radiation Imaging (Ministry of Education) and Department of Engineering Physics, Tsinghua University, Beijing 100084}
\author{H.~Y.~Xing}
\affiliation{College of Physics, Sichuan University, Chengdu 610065}
\author{R. Xu}
\affiliation{Key Laboratory of Particle and Radiation Imaging (Ministry of Education) and Department of Engineering Physics, Tsinghua University, Beijing 100084}
\author{Y.~Xu}
\affiliation{School of Physics, Nankai University, Tianjin 300071}
\author{T.~Xue}
\affiliation{Key Laboratory of Particle and Radiation Imaging (Ministry of Education) and Department of Engineering Physics, Tsinghua University, Beijing 100084}
\author{Y.~L.~Yan}
\affiliation{College of Physics, Sichuan University, Chengdu 610065}

\author{N.~Yi}
\affiliation{Key Laboratory of Particle and Radiation Imaging (Ministry of Education) and Department of Engineering Physics, Tsinghua University, Beijing 100084}
\author{C.~X.~Yu}
\affiliation{School of Physics, Nankai University, Tianjin 300071}
\author{H.~J.~Yu}
\affiliation{NUCTECH Company, Beijing 100084}
\author{J.~F.~Yue}
\affiliation{YaLong River Hydropower Development Company, Chengdu 610051}
\author{M.~Zeng}
\affiliation{Key Laboratory of Particle and Radiation Imaging (Ministry of Education) and Department of Engineering Physics, Tsinghua University, Beijing 100084}
\author{Z.~Zeng}
\affiliation{Key Laboratory of Particle and Radiation Imaging (Ministry of Education) and Department of Engineering Physics, Tsinghua University, Beijing 100084}
\author{B.~T.~Zhang}
\affiliation{Key Laboratory of Particle and Radiation Imaging (Ministry of Education) and Department of Engineering Physics, Tsinghua University, Beijing 100084}
\author{F.~S.~Zhang}
\affiliation{College of Nuclear Science and Technology, Beijing Normal University, Beijing 100875}
\author{L.~Zhang}
\affiliation{College of Physics, Sichuan University, Chengdu 610065}
\author{Z.~Y.~Zhang}
\affiliation{Key Laboratory of Particle and Radiation Imaging (Ministry of Education) and Department of Engineering Physics, Tsinghua University, Beijing 100084}
\author{K.~K.~Zhao}
\affiliation{College of Physics, Sichuan University, Chengdu 610065}
\author{M.~G.~Zhao}
\affiliation{School of Physics, Nankai University, Tianjin 300071}
\author{J.~F.~Zhou}
\affiliation{YaLong River Hydropower Development Company, Chengdu 610051}
\author{Z.~Y.~Zhou}
\affiliation{Department of Nuclear Physics, China Institute of Atomic Energy, Beijing 102413}
\author{J.~J.~Zhu}
\affiliation{College of Physics, Sichuan University, Chengdu 610065}

\collaboration{CDEX Collaboration}
\noaffiliation

\date{\today}

\begin{abstract}
We present novel constraints on boosted light dark matter particles (denoted as ``$\chi$'') from evaporating primordial black holes (PBHs) using 205.4 kg$\cdot$day data from the China Jinping Underground Laboratory's CDEX-10 p-type point contact germanium detector with a 160 eVee analysis threshold. $\chi$ from PBHs with masses ranging from 1$\times$10$^{15}$ g to 7$\times$10$^{16}$ g are searched in this work. In the presence of PBH abundance compatible with present bounds, our result excludes the $\chi$-nucleon elastic-scattering cross section region from 3.4$\times$10$^{-32}$ cm$^{2}$ to 2.3$\times$10$^{-29}$ cm$^{2}$ for $\chi$ of 1 keV to 24 MeV from PBHs with masses of 5$\times$10$^{15}$ g, as well as from 1.1$\times$10$^{-28}$ cm$^{2}$ to 7.6$\times$10$^{-28}$ cm$^{2}$ for $\chi$ of 1 keV to 0.6 MeV from PBHs with masses of 7$\times$10$^{16}$ g. If the $\chi$-nucleon elastic-scattering cross section can be determined in the future, the abundance of PBHs may be severely constrained by $\chi$ evaporation. With the lower threshold (160 eVee) of the CDEX-10 experiment compared to the previously used experiments, this work allows for a better reach at soft spectra produced by heavier PBHs, which demonstrates the vast potential of such a technical route to pursue $\chi$ from larger PBHs with a low threshold.
\end{abstract}

\maketitle

\section{Introduction}
Dark matter (DM) is an indispensable part of the standard cosmological model~\cite{RN1,RN2}. Some convincing evidence, such as galactic rotations, galaxy clusters, cosmic microwave backgrounds, and large-scale structures~\cite{RN3,RN4}, supports the existence of DM. However, light dark matter particles (denoted as ``$\chi$'') have not been definitely found to date in various experiments, such as collider searches and direct detection (DD) experiments~\cite{RN5}. DD experiments, such as XENON~\cite{RN6}, LUX-ZEPLIN~\cite{RN13}, PandaX~\cite{RN7}, DarkSide~\cite{RN10}, CRESST~\cite{RN9}, SuperCDMS~\cite{RN8}, CoGeNT~\cite{RN11}, and CDEX~\cite{RN14,RN15,RN16,RN17,RN18,RN19,RN20,RN21,cdexmidgal,RN22,RN36}, rely on $\chi$-nucleus ($\chi$-$N$) or $\chi$-electron scatterings through spin-independent (SI) and spin-dependent interactions. Collider searches provide upper bounds on the constraints of the $\chi$-nucleon cross section for lighter $\chi$~\cite{RN23}. However, in conventional searches with a traditional $\chi$ velocity distribution within the standard halo model~\cite{RN24,RN25}, DD experiments rapidly lose sensitivity toward the sub-GeV (sub-MeV) range because light $\chi$ carries insufficient energy to produce nuclear (electron) recoil signals with energies higher than the detector threshold. If $\chi$ can carry higher kinetic energy through specific mechanisms~\cite{RN26,RN27,RN28,PBHMiuDM,RN31,RN29,ABH_Cai}, a larger parameter space can be searched for lighter $\chi$. In recent years, it has been proposed that light $\chi$ can be upscattered or boosted to (semi)relativistic velocities through collisions with cosmic rays~\cite{RN31,RN29,RN30,RN32,RN33,RN34,RN35,RN36,CRDM_Wen_2019}, stellar neutrinos~\cite{stellar_neutrino1,stellar_neutrino2}, and diffuse supernova neutrinos~\cite{supernova_neutrino1,supernova_neutrino2}. Atmospheric collisions are also possible boosted $\chi$ sources ~\cite{PhysRevD.102.115032, PhysRevLett.123.261802, ARGUELLES2022137363}.

Primordial black holes (PBHs) are hypothetical black holes that could be formed soon after the inflationary epoch through the gravitational collapse of density fluctuations in the early Universe~\cite{RN37,RN38,PBHReviewMaxim_2010}; they could also be candidates for DM~\cite{RN57, RN74, PBHasDM_2014}. As a result of combining quantum field theory and general relativity, PBHs emit Hawking radiation, wherein the intensity corresponds to the PBH mass~\cite{RN39}. Based on experimental studies of Hawking radiation, stringent constraints on PBHs with masses of $M_{PBH}\sim\mathcal{O}$(10$^{14}$--10$^{17}$ g) have been obtained~\cite{RN40}. Hawking radiation comprises standard model particles such as photons, electrons, and neutrinos. $\chi$ can also be characterized as Hawking radiation~\cite{RN41,RN42,RN43,RN44,RN45,RN46,RN47,RN48,RN49,RN50}. Recently, a second potential strategy for locating boosted $\chi$~\cite{RN69,PBHDM_chi-e,PBHDM_Chie_Li} or new particle~\cite{PhysRevD.106.095034,ALPePBH_KA,AxionePBH_JY} from PBHs has been proposed. Some research has placed constraints on PBHs using 21 cm~\cite{RN51,Mittal_2022,PhysRevD.105.103026}, $\gamma$-rays~\cite{RN40,RN52,RN53,RN40,RN54}, positron-electrons~\cite{RN55,RN56,RN57}, neutrinos~\cite{RN58,RN59,Bernal_2022}, and dwarf galaxy gas heating~\cite{mnras_stab1222,LAHA2021136459}. If the $\chi$-$N$ cross section can be determined, then $\chi$ evaporation can be used to limit the fraction of PBHs.

In this study, the constraints on keV--MeV boosted $\chi$ from evaporating PBHs are derived based on the 205.4 kg$\cdot$day exposure data~\cite{cdex_darkphoton,RN36} from the CDEX-10 experiment, which aims at searching $\chi$ with p-type point contact germanium detectors~\cite{RN60} at the China Jinping Underground Laboratory (CJPL)~\cite{RN61}, where the physics analysis threshold is 160 eVee (electron equivalent energy).

\section{$\chi$ Spectra from Evaporating PBHs}
The Hawking temperature of an evaporating PBH with mass $M_{PBH}$ is ~\cite{RN62,RN63,RN64}
\begin{equation}
\begin{aligned}
T_{PBH} = \frac{\hbar c^3}{8\pi GM_{PBH} k_B},
\end{aligned}
\end{equation}
where $\hbar$ is the reduced Planck constant, $c$ is the speed of light in vacuum, $G$ is the Newtonian constant of gravitation, and $k_B$ is the Boltzmann constant. PBHs with larger masses have lower Hawking temperatures. The differential number of $\chi$ per unit time evaporated by PBHs is calculated by the public code $\tt Blackhawk$-$\tt v2.1$~\cite{RN65,RN66}
\begin{equation}
\begin{aligned}
\frac{d^2N_\chi }{dE_\chi dt} = \frac{g_\chi }{2\pi } \frac{\Gamma_\chi (E_\chi,M_{PBH})}{exp(E_\chi/k_BM_{PBH})+1},
\end{aligned}
\end{equation}
where $E_\chi$ is the evaporated $\chi$ energy and $\Gamma_\chi$ is the graybody factor. Here, we consider the case of Dirac fermions with four degrees of freedom ($g_\chi$=4) and chargeless and spinless PBHs. 

\section{$\chi$ Flux Reaching Earth}
For a monochromatic PBH mass distribution, the flux of $\chi$ reaching Earth consists of a galactic (MW) and an extragalactic (EG) component,
\begin{equation}
\begin{aligned}
\frac{d^2\phi_\chi}{dT_\chi d\Omega} = \frac{d^2\phi_\chi^{MW}}{dT_\chi d\Omega}+\frac{d^2\phi_\chi^{EG}}{dT_\chi d\Omega},
\end{aligned}
\label{eq::eq3}
\end{equation}
where $T_\chi$ is the kinetic energy of $\chi$. The MW component is
\begin{equation}
\begin{aligned}
\frac{d^2\phi_\chi^{MW}}{dT_\chi d\Omega } = \frac{1}{4\pi }\frac{f_{PBH}}{M_{PBH}} \int \frac{d\Omega_s}{4\pi}\int dl \rho _{MW}^{\rm NFW}[r(s,\phi)]\frac{d^2N_\chi}{dT_\chi dt},
\end{aligned}
\end{equation}
where $\rho_{MW}^{\rm NFW}$ is the DM density of the Milky Way halo in the Navarro-Frenk-White DM profile~\cite{RN67} with a local density of $\rho_ \odot$ = 0.4 GeV/cm$^{3}$, $r(s,\phi)$ is the galactocentric distance with the solar distance from the galactic center, $s$ is the line-of-sight distance to the PBH, and $\phi$ is the angle between these two directions. $f_{PBH}$ is the fraction of DM composed of PBHs. For PBHs with masses ranging from $1 \times 10^{15}$ g to $7 \times 10^{16}$ g, the strictest constraints on PBHs are from EDGES 21 cm~\cite{RN51} and COMPTEL~\cite{RN54}. The $f_{PBH}$ values chosen in this work are listed in Table~\ref{tab:addlabel} and marked with black circles in Fig.~\ref{fig::fig6}. An analysis cutoff is applied to $M_{PBH} = 7\times10^{16}$ g since the existing experimental constraints on $f_{PBH}$ at larger $M_{PBH}$ are weak.
\begin{table}[!htbp]
  \centering
  \caption{Selected $f_{PBH}$ values at different $M_{PBH}$s for this work, based
on the latest constraints from Refs.~\cite{RN51},~\cite{RN54}.}
    \begin{tabular*}{\hsize}{@{\quad}@{\extracolsep{\fill}}cccc@{\quad}}
    \hline
    \hline
    $M_{PBH}$ (g) & $f_{PBH}$ & $M_{PBH}$ (g) & $f_{PBH}$ \\
    \hline
    $1\times 10^{15}$ & $1.6\times 10^{-8}$ & $9\times 10^{15}$ & $5.2\times 10^{-5}$ \\
    $2\times 10^{15}$ & $1.8\times 10^{-7}$ & $1\times 10^{16}$ & $7.0\times 10^{-5}$ \\
    $3\times 10^{15}$ & $9.8\times 10^{-7}$ & $2\times 10^{16}$ & $1.0\times 10^{-3}$ \\
    $4\times 10^{15}$ & $3.5\times 10^{-6}$ & $3\times 10^{16}$ & $3.8\times 10^{-3}$ \\
    $5\times 10^{15}$ & $8.6\times 10^{-6}$ & $4\times 10^{16}$ & $7.6\times 10^{-3}$ \\
    $6\times 10^{15}$ & $1.4\times 10^{-5}$ & $5\times 10^{16}$ & $1.4\times 10^{-2}$ \\
    $7\times 10^{15}$ & $2.0\times 10^{-5}$ & $6\times 10^{16}$ & $2.2\times 10^{-2}$ \\
    $8\times 10^{15}$ & $3.5\times 10^{-5}$ & $7\times 10^{16}$ & $3.4\times 10^{-2}$ \\
    \hline
    \hline
    \end{tabular*}
    \raggedright
     Note: ($M_{PBH}$ ($10^{15}$ g), $f_{PBH}$)s used by Calabrese $et\ al.$ in Ref.~\cite{RN69} are (0.5, $2.9\times10^{-10}$), (1.0, $3.9\times10^{-7}$) and (8.0, $3.7\times10^{-4}$).
  \label{tab:addlabel}
\end{table}

For the EG component in Eq.~\ref{eq::eq3}, the effect of redshift $z(t)$ on the kinetic energy should be taken into account,
\begin{equation}
\begin{aligned}
\frac{d^2\phi_\chi^{EG}}{dT_\chi d\Omega } =\frac{f_{PBH}\rho _{DM}}{4\pi M_{PBH}} \int dt [1+z(t)] \frac{d^2N_\chi }{dT_\chi dt},
\end{aligned}
\label{eq::eq5}
\end{equation}
where $\rho_{DM}$ = 2.35$\times $10$^{-30}$ g/cm$^3$ is the average DM density of the Universe at the current epoch~\cite{RN68}. The integral in Eq.~\ref{eq::eq5} is from the time of matter-radiation equality to the age of the Universe.

\begin{figure}[!htbp]
\includegraphics[width=0.99\linewidth]{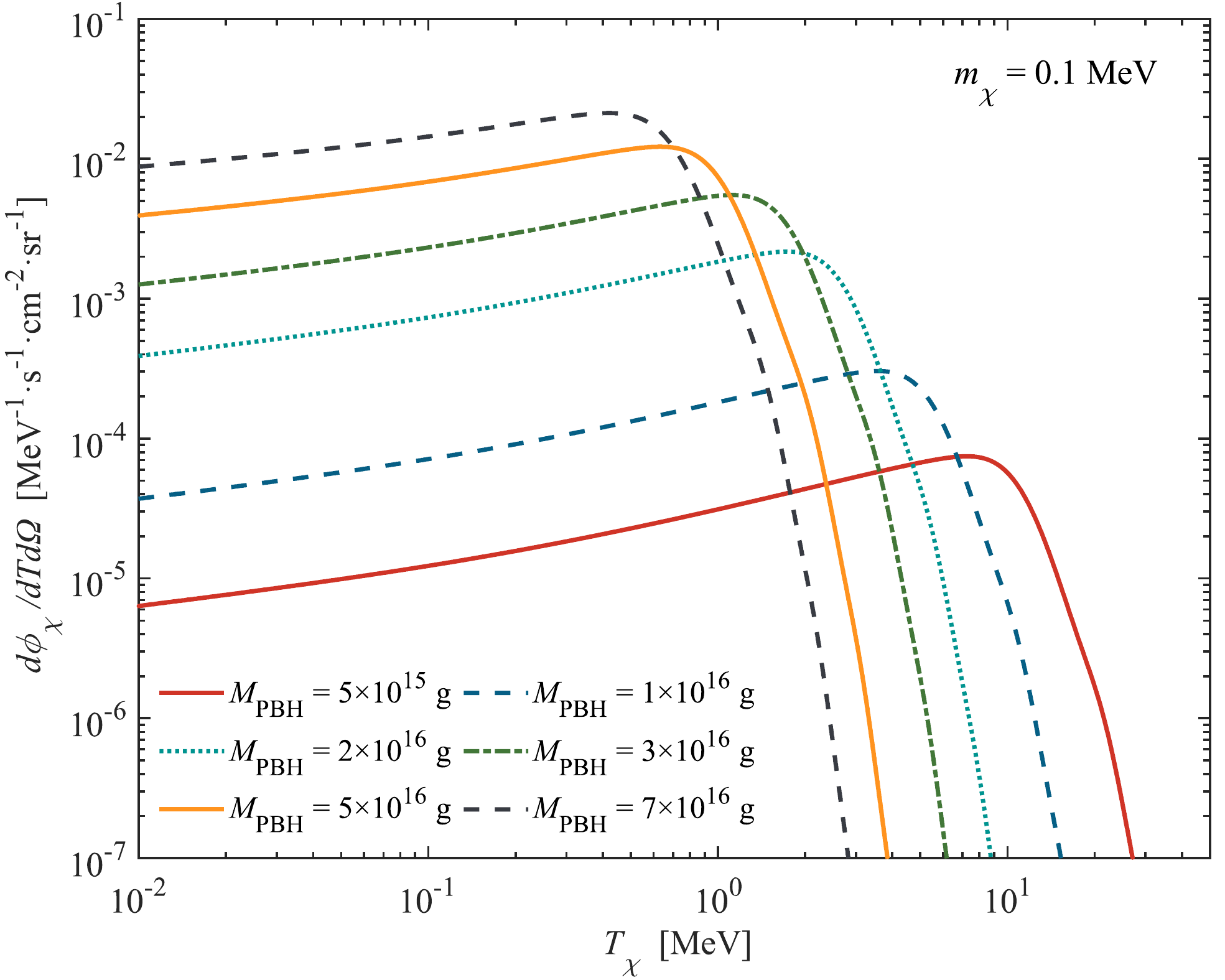}
\caption{
The fluxes of $\chi$ with mass $m_\chi$ = 0.1 MeV reaching Earth from PBHs with different masses. The fraction of DM comprising PBHs chosen in our calculation is listed in Table~\ref{tab:addlabel}.
}
\label{fig::fig1}
\end{figure}

Figure~\ref{fig::fig1} shows the flux of $\chi$ with mass $m_\chi$ = 0.1 MeV reaching Earth from PBHs. PBHs with larger masses emit softer $\chi$ spectra.

\section{Earth attenuation}
$\chi$ must travel through a few kilometers of rock before reaching the underground laboratory. At large scattering cross sections with a mean free path in rocks less than a few kilometers, $\chi$ will scatter with the nucleus and lose kinetic energy before reaching the detectors at underground locations. This phenomenon is called the Earth shielding effect. In this work, this process is modeled with a ballistic-trajectory approximation following Refs.~\cite{RN31,RN69}, which is subjected to systematic uncertainties in the numerical calculations~\cite{RN36,ESS_Liu}. We note that different results are obtained by different numerical frameworks~\cite{RN29,RN36,Xia_2022}. The subject of $\chi$ attenuation by the Earth remains an open research question and will be a theme of our future efforts.

The $\chi$ flux will be changed from $\frac{d^2\phi_\chi}{dT_0 d\Omega }$ to $\frac{d^2\phi_\chi^d}{dT_d d\Omega }$, where $d$ represents the depth of rock through which incident $\chi$ passes, $T_0$ represents the initial kinetic energy of $\chi$, and $T_d$ represents the kinetic energy of $\chi$ reaching the detectors after traveling a distance of $d$. The $\chi$ flux arriving at the CJPL can be evaluated by
\begin{equation}
\begin{aligned}
\frac{d^2\phi_\chi^d}{dT_d d\Omega } \approx \frac{4 m_\chi^2e^\tau}{(2 m_\chi+T_d-T_de^\tau)^2} \left(\frac{d^2\phi_\chi}{dT_0 d\Omega }\right),
\end{aligned}
\label{eq::eq6}
\end{equation}
with
\begin{equation}
\begin{aligned}
T_d(T_0) = \frac{2 m_\chi T_0}{2 m_\chi e^\tau - T_0(1+e^\tau)},
\end{aligned}
\end{equation}
where $\tau = d/l$, $d$ = 2400 m is the rock overburden depth of the CJPL~\cite{RN61} and $l$ is the interaction length given by
\begin{equation}
\begin{aligned}
l=\left[ \sum_N n_N \sigma_{\chi N} \frac{2 m_N m_\chi}{(m_N+m_\chi)^2} \right]^{-1},
\end{aligned}
\label{eq::eq7}
\end{equation}
where $n_N$ is the number density of element $N$. The density and the elemental compositions of the rocks are provided in Ref.~\cite{ESS_Liu}. Omitting the nuclear form factor, the $\chi$-nucleus elastic scattering cross section $\sigma_{\chi N}$ can be expressed with
\begin{equation}
\begin{aligned}
\sigma_{\chi N} = \sigma^{\rm SI}_{\chi p} A_N^2 \left[\frac{m_N(m_\chi +m_p)}{m_p(m_\chi +m_N)}\right]^2,
\end{aligned}
\label{eq::eq8}
\end{equation}
where $A_N$ is the mass number of nucleus $N$, $\sigma_{\chi p}^{\rm SI}$ is the SI $\chi$-nucleon elastic scattering cross section, and $m_N$ and $m_p$ are the masses of nucleus $N$ and proton, respectively. This approximation of omitting the form factor will result in an error of less than 1\% because the nuclear recoil energy spectra are softer than the ones in Ref.~\cite{ESS_Liu}.

\section{Expected event rate in Ge detectors}
The differential event rate of the $\chi$-germanium nucleus scattering in the detector is calculated by
\begin{equation}
\begin{aligned}
\frac{dR}{dE_r}(E_r) = \sigma_{\chi Ge}N_{Ge}\int dT_dd\Omega \frac{d^2\phi_\chi^d}{dT_d d\Omega} \frac{\Theta(E_r^{max}-E_r)}{E_r^{max}},
\end{aligned}
\label{eq::eq10}
\end{equation}
where $N_{Ge}$ = 8.30$\times$10$^{21}$ g$^{-1}$ is the number of germanium nuclei per gram and $\Theta$ is the Heaviside function. $E_r^{max}$ is the maximum value of the recoil energy $E_r$ expressed with 
\begin{equation}
\begin{aligned}
E_r^{max} = \frac{T_d^2+2 m_\chi T_d}{T_d+(m_\chi +m_{Ge})^2/(2 m_{Ge})}.
\end{aligned}
\label{eq::eq12}
\end{equation}

The $\chi$-germanium nucleus elastic scattering differential cross section $\sigma_{\chi Ge}$ is given by
\begin{equation}
\begin{aligned}
\sigma_{\chi Ge} = F^2(E_r)\sigma_{\chi p}^{\rm SI}A_{Ge}^2\left[\frac{m_{Ge}(m_\chi+m_p)}{m_p(m_\chi+m_{Ge})}\right]^2, 
\end{aligned}
\label{eq::eq11}
\end{equation}
where $F(E_r)$ is the nuclear form factor, for which the conventional Helm form factor~\cite{RN71} is adopted. $\sigma_{\chi p}^{\rm SI}$ is the SI $\chi$-nucleon elastic scattering cross section. $m_p$ is the mass of the proton. $A_{Ge} = 72.5$ is the mass number of germanium, and $m_{Ge}$ = 6.77$\times$10$^4$ MeV is the mass of germanium nuclei.

In a germanium semiconductor detector, the observed total deposit energy $E_{det}$ is different from the real nuclear-recoil energy $E_r$ and should be corrected by the quenching factor, $E_{det}=Q_{nr}E_r$. The differential event rate per deposit energy $E_{det}$ can be obtained as follows:
\begin{equation}
\begin{aligned}
\frac{dR}{dE_{det}} = \frac{dR}{dE_r}\cdot \left(\frac{dQ_{nr}}{dE_r}\cdot E_r+Q_{nr}\right)^{-1}, 
\end{aligned}
\label{eq::eq13}
\end{equation}
where the quenching factor $Q_{nr}$ is determined using the Lindhard formula~\cite{RN72}, and $\kappa$ = 0.16 (a typical value reported in the literature that closely matches recent measurements in the low-energy range~\cite{quenching2022}).

\begin{figure}[!htbp]
\includegraphics[width=0.96\linewidth]{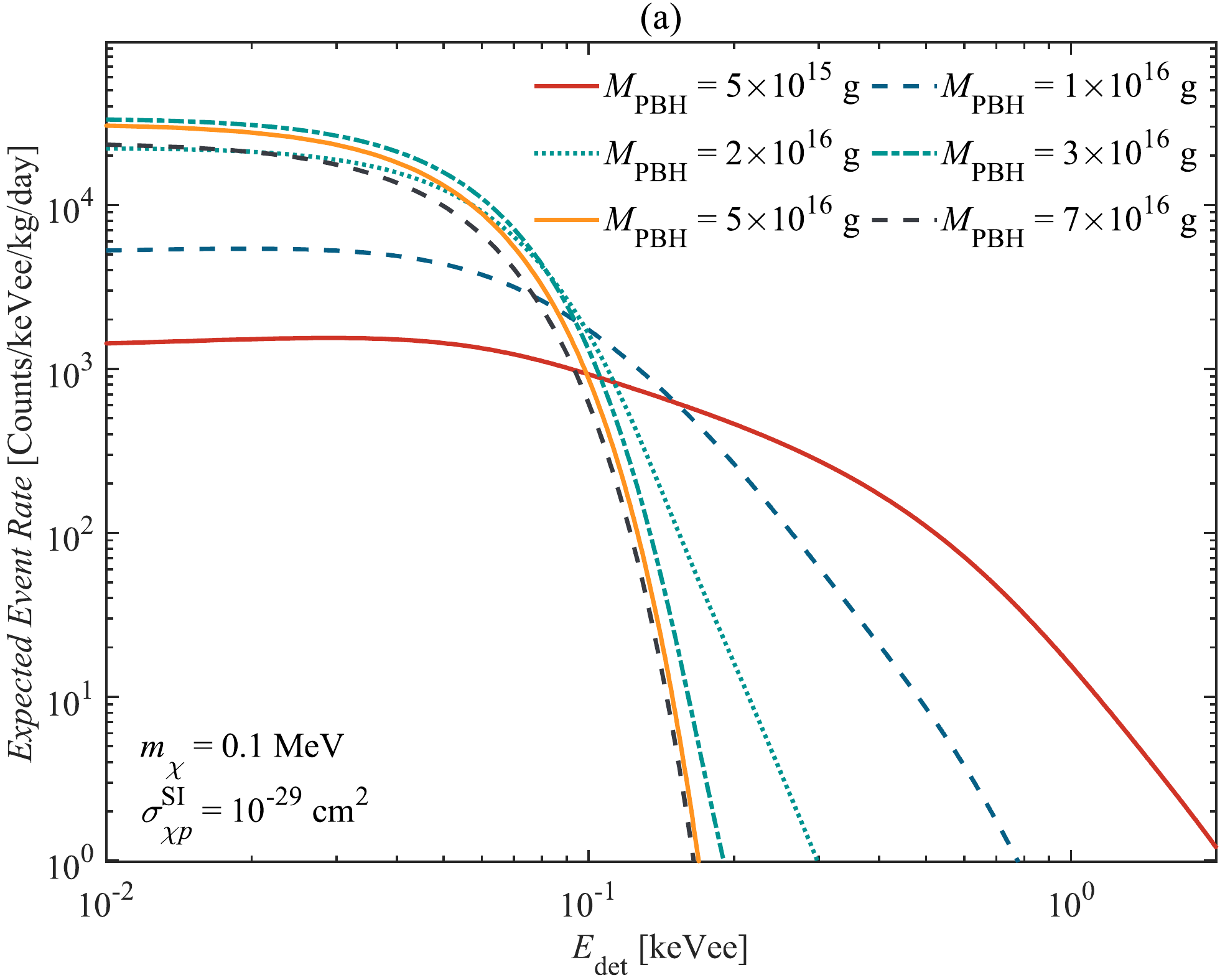}
\includegraphics[width=0.96\linewidth]{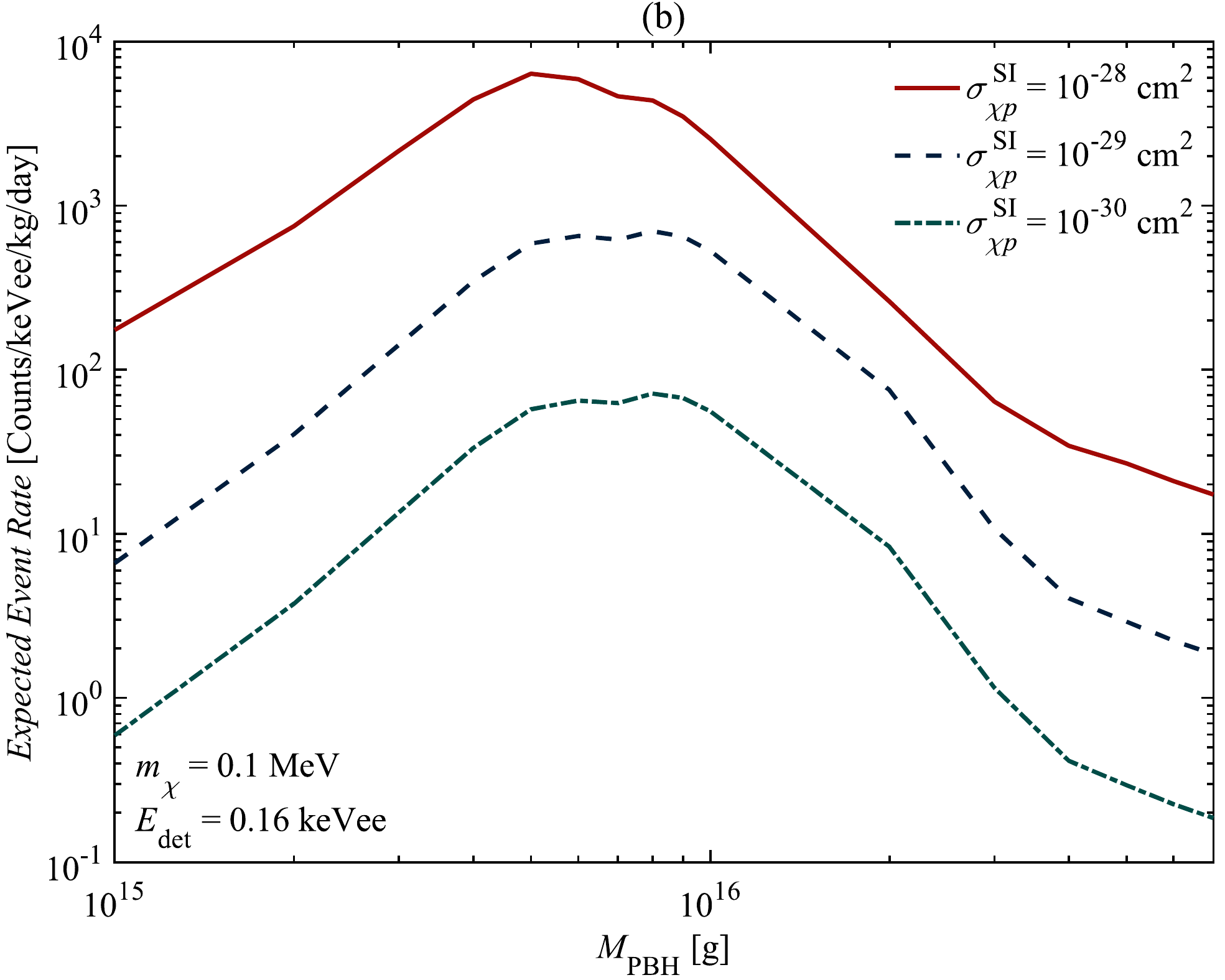}
\caption{
(a) Expected event rates for different $M_{PBH}$s when $m_\chi$ = 0.1 MeV and $\sigma^{\rm SI}_{\chi p}$ = 10$^{-29}$ cm$^2$. (b) Expected event rates at 0.16 keVee in the CDEX-10 detector for $\sigma_{\chi p}^{\rm SI}$ = 10$^{-28}$ cm$^2$, 10$^{-29}$ cm$^2$, and 10$^{-30}$ cm$^2$. The quenching factor in germanium is determined by the Lindhard formula~\cite{RN72} and $\kappa $ = 0.16~\cite{quenching2022}. The energy resolution of CDEX-10 was considered, and its standard deviation is 35.8 + 16.6$\times E^{\frac {1}{2}}$ (eV), where $E$ is expressed in keV~\cite{RN20,RN36,CDEX_DM_e}.
}
\label{fig::fig2}
\end{figure}

The expected event rates in a germanium detector convolved with energy resolution for different $M_{PBH}$s are shown in Fig.~\ref{fig::fig2}(a) when $m_\chi$ = 0.1 MeV and $\sigma^{\rm SI}_{\chi p}$ = 10$^{-29}$ cm$^2$. PBHs with larger masses have softer expected energy spectra. As shown in Fig.~\ref{fig::fig2}(b), the expected event rate decreases at $E_{det}$ = 0.16 keVee, whereas the $\chi$ spectrum becomes softer as the PBH mass increases when $M_{PBH} \gtrsim 8\times 10^{15}$ g. The search for $\chi$ from PBHs with larger masses is more sensitive to the threshold.

Figure~\ref{fig::fig3} shows the expected event rates for boosted $\chi$ from evaporating PBHs at the specified parameters ($M_{PBH}$, $\sigma_{\chi p}^{\rm SI}$, $B$). As $M_{PBH}$ becomes larger, the expected spectrum becomes significantly softer. This means that low-threshold detectors can perform well in the search for $\chi$ from PBHs with larger masses. 

\section{Exclusion result and discussion}
\begin{figure}[!tbp]
\includegraphics[width=\linewidth]{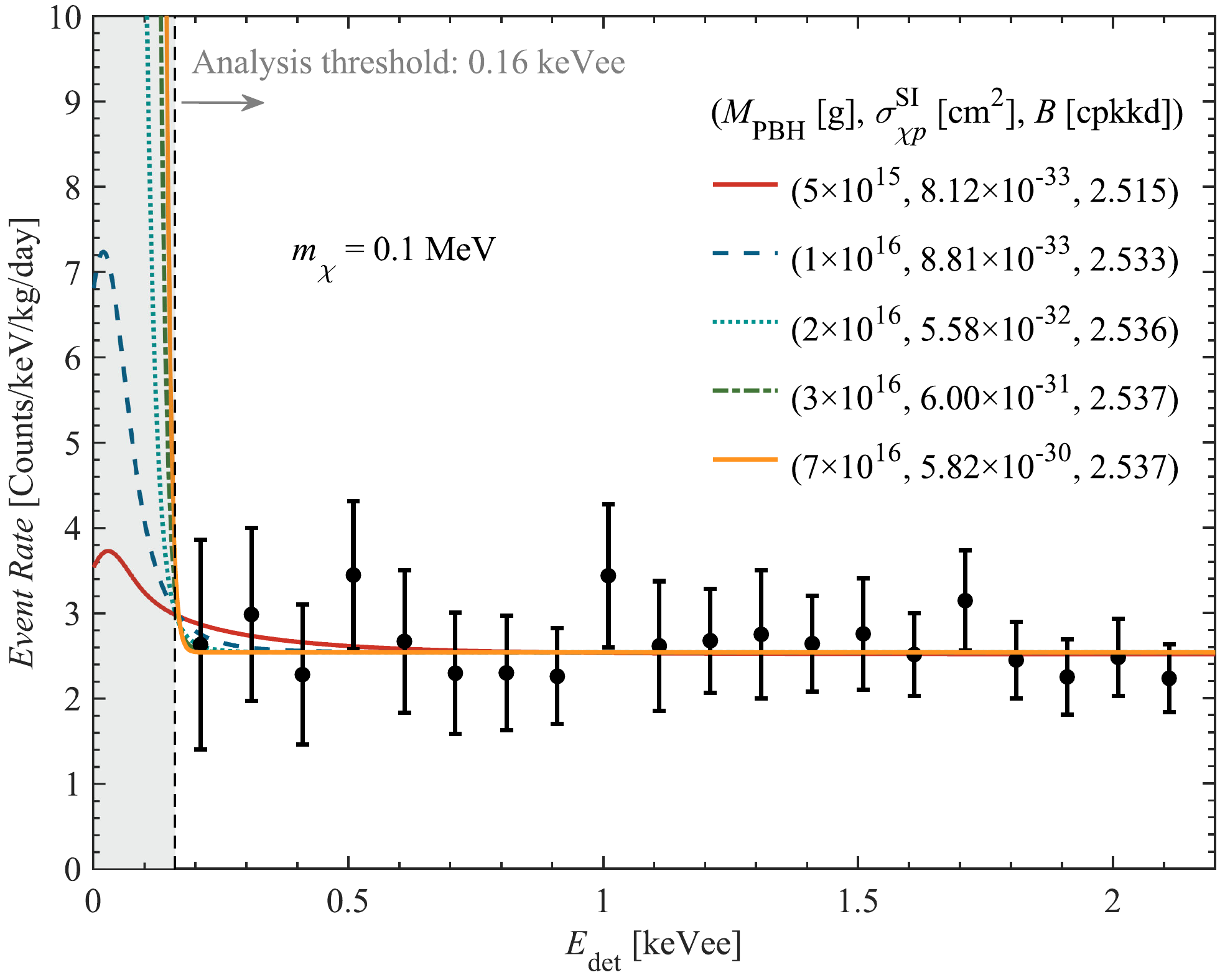}
\caption{
Expected event rates (shown by lines) in the CDEX-10 detector for light $\chi$ from evaporating PBHs at the specified parameters ($M_{PBH}$, $\sigma_{\chi p}^{\rm SI}$, $B$) chosen from the best fit of the minimum-$\chi^2$. ``cpkkd'' means ``Counts/keVee/kg/day''. The energy resolution of CDEX-10 was considered. The measured spectrum of CDEX-10 is shown by black points with error bars, which shows the residual spectrum with the L- and M-shell x-ray contributions subtracted in the region of 0.16$-$2.16 keVee, with a bin width of 100 eVee~\cite{RN36,CDEX_DM_e}.
}
\label{fig::fig3}
\end{figure}

Data analysis for this work is based on a 205.4 kg$\cdot$day dataset from CDEX-10. At the subkeVee energy range relevant to this analysis, background events are dominated by the Compton scattering of high-energy gamma rays and internal radioactivity from long-lived cosmogenic isotopes. Figure~\ref{fig::fig3} shows the measured spectrum after subtracting the contributions from the L- and M-shell x-ray peaks derived from the corresponding K-shell x-ray intensities. Following our previous WIMP, dark photon, and $\chi$-electron scattering analysis~\cite{RN19,cdex_darkphoton,CDEX_DM_e}, a minimum-$\chi^2$ analysis is applied to the residual spectrum in the range of 0.16$-$2.16 keVee:
\begin{equation}
\begin{aligned}
& \chi^2 \left(M_{PBH}, f_{PBH}, m_{\chi}, \sigma^{\rm SI}_{\chi p}\right) = \\
& \sum_i \frac{\left[n_i-B-S_i\left(M_{PBH}, f_{PBH}, m_{\chi}, \sigma^{\rm SI}_{\chi p}\right)\right]^2}{\Delta^2_i},
\end{aligned}
\end{equation}
where $n_i$ and $\Delta _i$ denote the measured data and standard deviation with statistical and systematic components at the $i$th energy bin, respectively; $S_i$ denotes the predicted event rate; and $B$ denotes the assumed flat-background contribution from the Compton scattering of high-energy gamma rays.

\begin{figure}[!tbp]
\includegraphics[width=\linewidth]{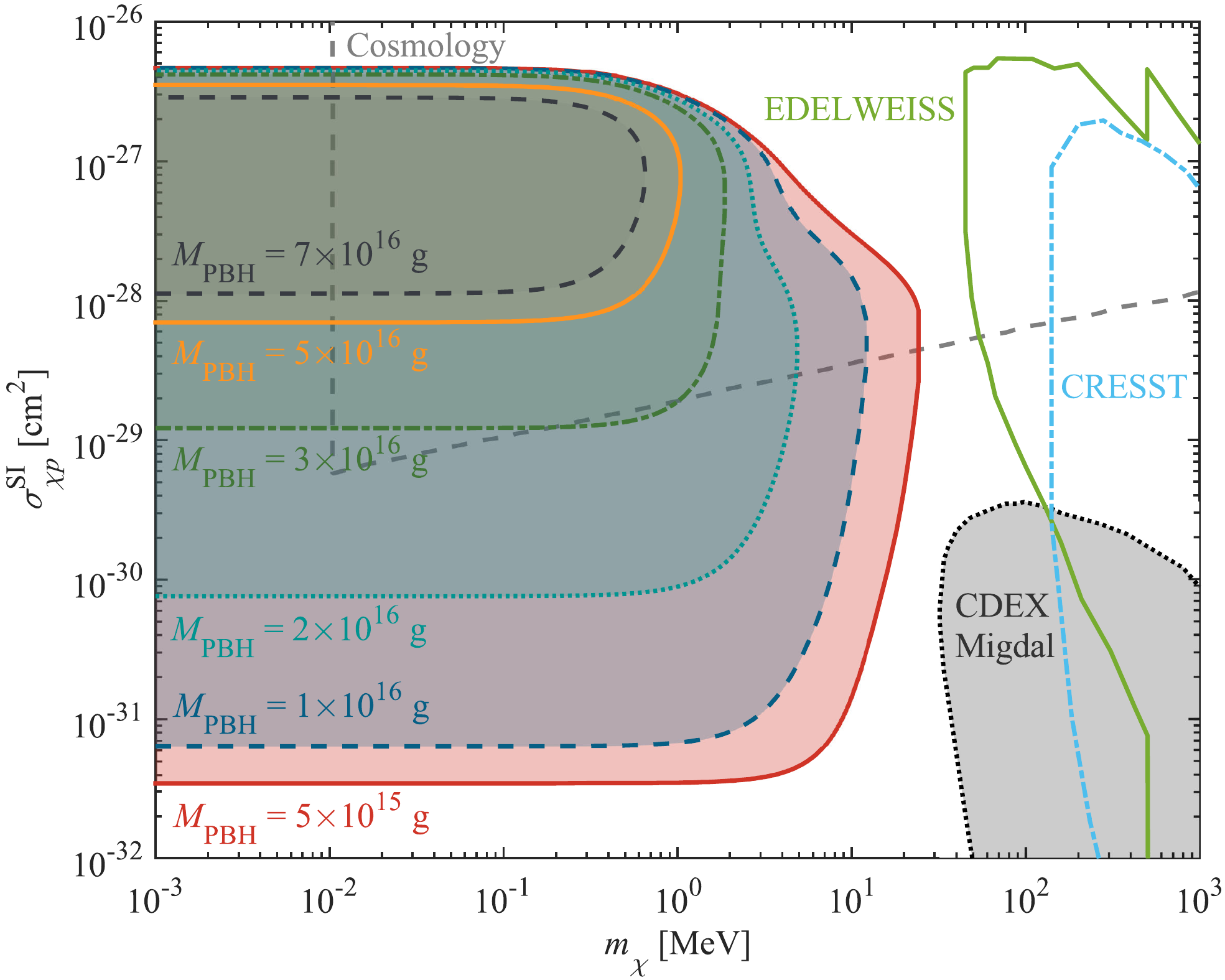}
\caption{Exclusion regions ($m_\chi$, $\sigma_{\chi p}^{\rm SI}$) with 90\% confidence level for $M_{PBH}$ = (0.5, 1, 2, 3, 5, 7) $\times$10$^{16}$ g. The exclusion region from cosmology~\cite{PhysRevLett.121.081301, PhysRevD.97.103530, PhysRevD.98.023013, Nadler_2019} and the published limits under the standard halo model scenario from CDEX-10 Migdal effect analysis~\cite{ESS_Liu}, CRESST $\nu$-cleus 2017 surface run~\cite{CREEST_ES,CREEST_WIMP,CRESST_MeV}, and EDELWEISS-surface~\cite{EDELWEISS} are superimposed. For visual clarity, the constraints from phenomenological interpretations of the XENON-1T data~\cite{RN69} are omitted from this figure.}
\label{fig::fig4}
\end{figure}

Figure~\ref{fig::fig4} shows the 90\% confidence level (C.L.) exclusion regions ($m_\chi$, $\sigma_{\chi p}^{\rm SI}$) for $M_{PBH}$ = 0.5, 1, 2, 3 and 7$\times$10$^{16}$ g, derived by the Feldman-Cousins unified approach with $\Delta \chi^2$ = 1.64~\cite{RN73}. The constraints from the Migdal effect analysis of CDEX-10 data~\cite{ESS_Liu} and cosmology~\cite{PhysRevLett.121.081301, PhysRevD.97.103530, PhysRevD.98.023013, Nadler_2019}, CRESST $\nu$-cleus 2017 surface run~\cite{CREEST_ES,CREEST_WIMP,CRESST_MeV}, and EDELWEISS-Surface~\cite{EDELWEISS} are shown for comparison. For visual clarity, the constraints from phenomenological interpretations of the XENON-1T data~\cite{RN69}, obtained for different values of $M_{PBH}$ = (0.5, 1.0, 8.0) $\times 10^{15}$ g, are omitted from the figure.

\begin{figure}[!tbp]
\includegraphics[width=\linewidth]{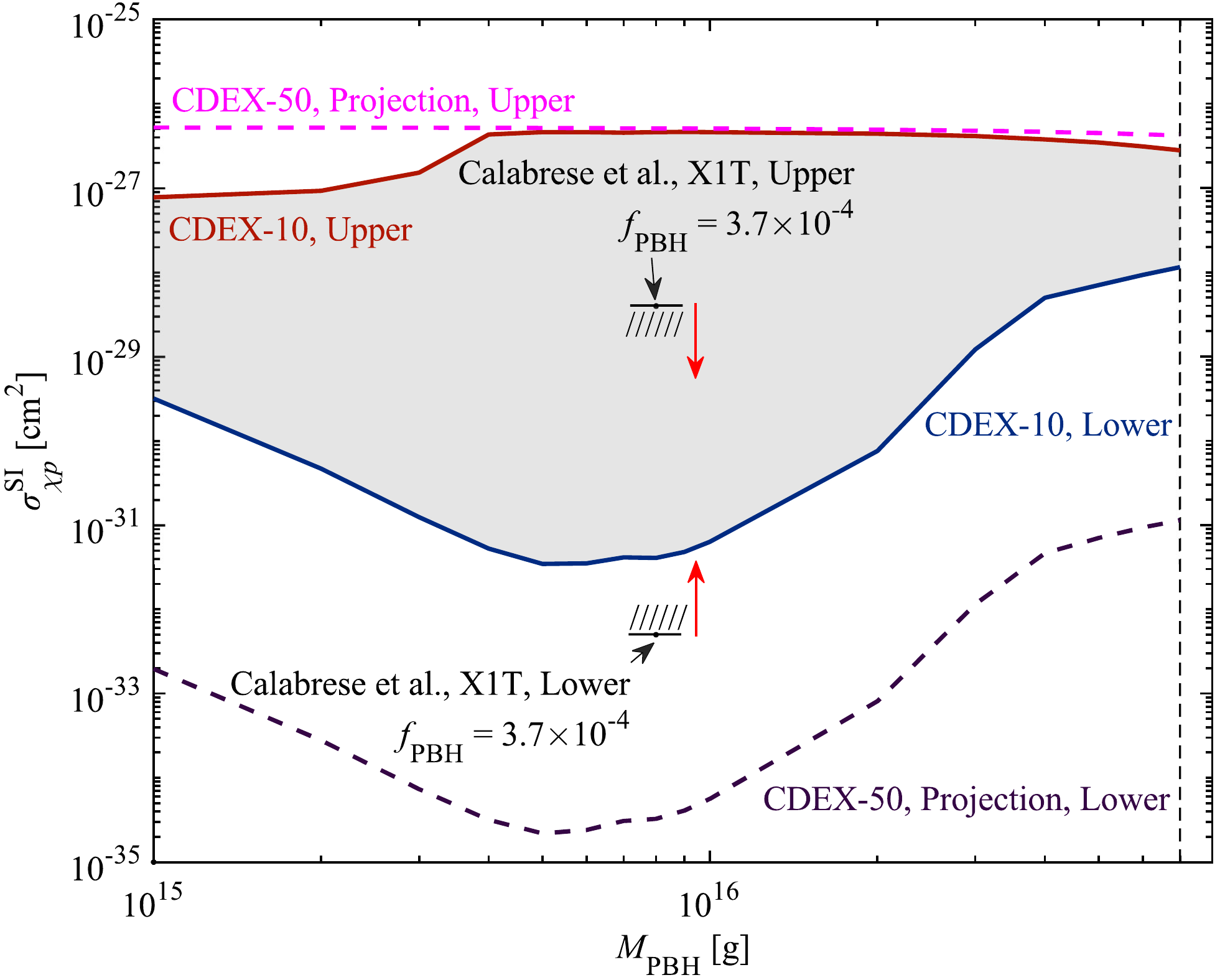}
\caption{Exclusion regions ($M_{PBH}$, $\sigma_{\chi p}^{\rm SI}$) with 90\% C.L. for $m_\chi$ $\lesssim$ 0.1 MeV. The lower and upper constraints on $\sigma_{\chi p}^{\rm SI}$ are shown by blue and orange solid lines, respectively. The gray-shaded parameter space in between is probed and excluded.  The projected sensitivities from CDEX-50 are shown by dotted dashed lines. An analysis cutoff is applied to $M_{PBH}=7\times10^{16}$ g since the existing experimental constraints on $f_{PBH}$ at larger $M_{PBH}$ are weak. The constraints from phenomenological interpretations by Calabrese $et\ al.$ of the XENON-1T data~\cite{RN69} for $M_{PBH}$ = 8$\times10^{15}$ g are also plotted in horizontal lines with one-side shading. If the latest $f_{PBH}$ values in Table~\ref{tab:addlabel} are adopted, the excluded parameter space will be reduced, as indicated by the red arrows.}
\label{fig::fig5}
\end{figure}

As Fig.~\ref{fig::fig4} shows, both the upper and lower lines of the exclusion regions for $\chi$ with masses below 0.1 MeV are flat. We search the exclusion regions for $M_{PBH}$ from 1$\times$10$^{15}$ g to 7$\times$10$^{16}$ g choosing $m_\chi$ = 0.1 MeV as a benchmark value for our analysis. The results are shown in Fig.~\ref{fig::fig5}. The constraints from phenomenological interpretations by Calabrese $et\ al.$ of the XENON-1T data~\cite{RN69} for $M_{PBH}$ = 8$\times10^{15}$ g are also plotted. Adoption of the latest $f_{PBH}$ values in Table~\ref{tab:addlabel} would give rise to 
weaker constraints as indicated by the red arrows, such that this work significantly extends the excluded parameter space.

The next generation of the CDEX experiment, CDEX-50, is currently under construction. A 50 kg germanium detector array will be run in a low-radioactive environment, and the radioactive background will be further reduced to $\sim$ 0.01 cpkkd in the sub-keV region~\cite{CDEX_DM_e}. As shown by the purple dashed line in Fig.~\ref{fig::fig5}, the lower constraints on $\sigma_{\chi p}^{\rm SI}$ from 50 kg$\cdot$year data of CDEX-50 will be improved by up to approximately three orders of magnitude compared to those from CDEX-10. At the same time, $\chi$ from PBHs with a broader mass range can be effectively searched.

\begin{figure}[!htbp]
\includegraphics[width=\linewidth]{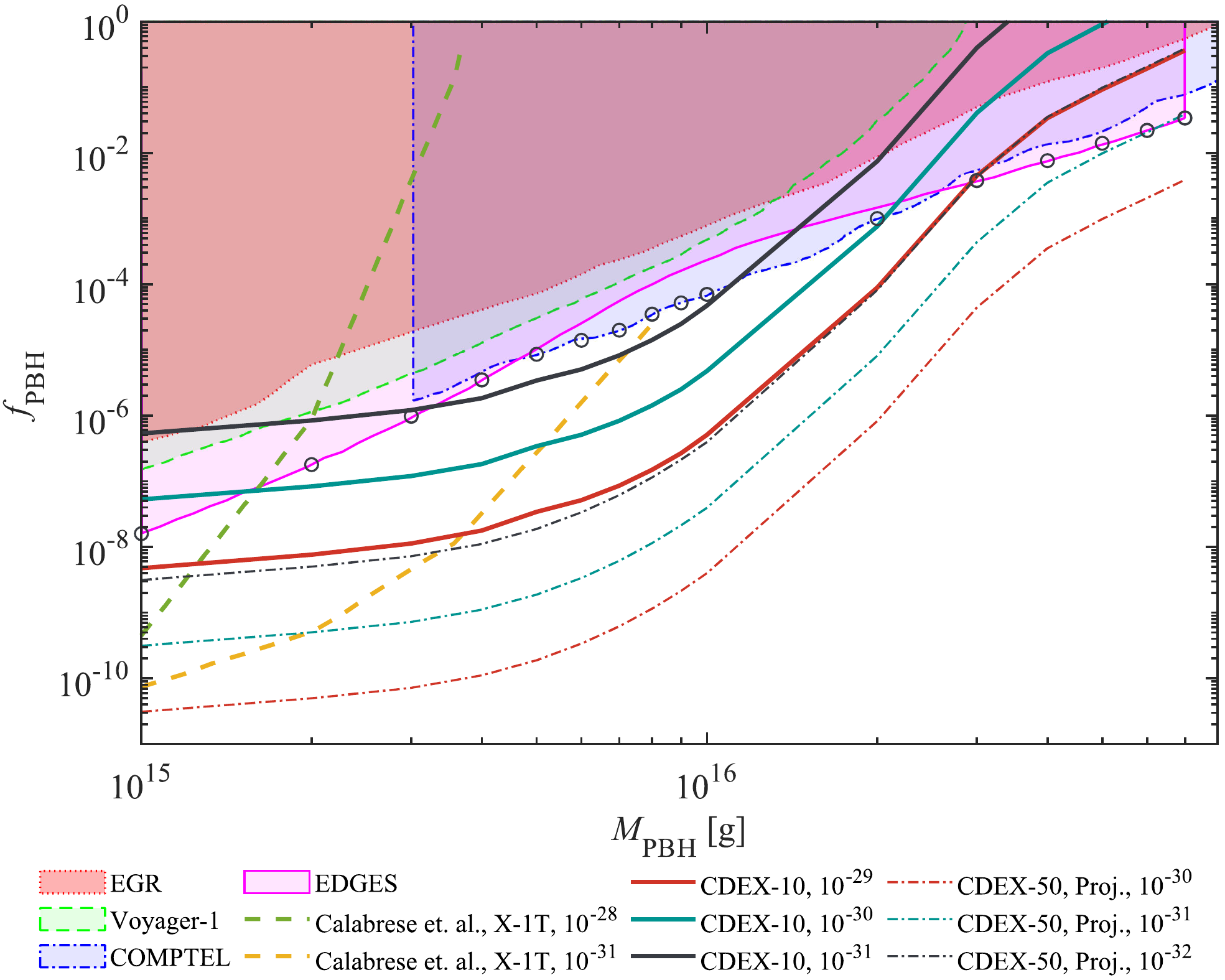}
\caption{
Upper bounds with 90\% C.L. on $f_{PBH}$ were obtained for different $\sigma_{\chi p}^{\rm SI}$s ($m_\chi$ $\lesssim$ 0.1 MeV). Some strong existing bounds from extragalactic $\gamma$-ray background (EGR)~\cite{RN53}, EDGES 21 cm~\cite{RN51}, Voyager-1~\cite{RN55}, COMPTEL~\cite{RN54} observations are shown, and the strongest ones marked with black circles are selected for input $f_{PBH}$ listed in Table~\ref{tab:addlabel}. Projection results from CDEX-50 are shown by thin dotted-dashed lines. The numbers in the legend, 10$^{-29}$, 10$^{-30}$, 10$^{-31}$ and 10$^{-32}$, are $\sigma_{\chi p}^{\rm SI}$s in the unit of cm$^2$. The constraints from phenomenological interpretations by Calabrese $et\ al.$ of the XENON-1T data are also shown by dashed lines, where for $\sigma^{\rm SI}_{\chi p}$ = 10$^{-28}$ cm$^2$, the Earth shielding effect is more evident, pushing the events to lower-recoil energies and weakening the XENON-1T's sensitivity~\cite{RN69}.
}
\label{fig::fig6}
\end{figure}

In other words, if the Earth shielding effect can be omitted; as $f_{PBH}$ decreases, $\sigma_{\chi p}^{\rm SI}$ decreases qualitatively. We fix $\sigma_{\chi p}^{\rm SI}$ = 10$^{-29}$, 10$^{-30}$ and 10$^{-31}$ cm$^2$ and then search for $f_{PBH}$ with $m_\chi$ = 0.1 MeV. Figure~\ref{fig::fig6} shows the upper bounds with 90\% C.L. on $f_{PBH}$, obtained for different $\sigma_{\chi p}^{\rm SI}$. According to the projection results (thin dotted dashed lines in Fig.~\ref{fig::fig6}), CDEX-50 can be well placed to search for $\chi$ from PBHs with a mass of $M_{PBH}$ $\gtrsim$ 2$\times$10$^{16}$ g. Some strong existing bounds from extragalactic $\gamma$-ray background (EGR)~\cite{RN53}, EDGES 21 cm~\cite{RN51}, Voyager-1~\cite{RN55}, and COMPTEL~\cite{RN54} observations are superimposed. The constraints from phenomenological interpretations by Calabrese $et\ al.$ of the XENON-1T data~\cite{RN69} are also shown by dashed lines for comparison.

Finally, there is the degeneracy of $\sigma_{\chi p}^{\rm SI}$ and $f_{PBH}$ to consider. Following the determination of one of $\sigma_{\chi p}^{\rm SI}$ and $f_{PBH}$ in the future measure, the other one can be known. For example, if PBHs exist and $f_{PBH}$ is accurately measured in the observation of Hawking radiation, such as $\gamma$-rays and neutrinos, accurate constraints on $\sigma_{\chi p}^{\rm SI}$ can be derived. As the spectrum becomes softer, the CDEX experiment has vast potential as a technical route to pursue $\chi$ from larger PBHs with a low threshold.

\acknowledgments
We would like to thank Marco Chianese, Tong Li, and Wei Chao for their useful discussions. This work was supported by the National Key Research and Development Program of China (Grants No. 2017YFA0402200 and No. 2022YFA1605000) and the National Natural Science Foundation of China (Grants No. 12175112, No. 12005111, and No. 11725522).

\bibliography{ePBH_DM.bib}

\end{document}